# Effects of the Two- Gap Nature on the Microwave Conductivity of 39 K Polycrystalline MgB$_2$ Films

Sang Young Lee[1,*], J. H. Lee[1], Jung Hoon Han[2,3], S. H. Moon[4], H. N. Lee[5], James C. Booth[6] and J. H. Claassen[7]

*1 Department of Physics and Center for Optoelectronic and Microwave Devices,
Konkuk University, Seoul 143-701, Korea
2 Department of Physics, Sungkyunkwan University, Suwon, Korea,
3 CSCMR, Seoul National University, Seoul 151-747, Korea
4 School of Materials Science, Seoul National University, Seoul, Korea
5 LG Electronic Institute of Technology, Seoul, Korea
6 National Institute of Standards and Technology, Boulder, CO, U.S.A.
7 Naval Research Laboratory, Washington D.C., U.S.A.*

The surface resistance ($R_S$) and the real part ($\sigma_1$) of the microwave complex conductivity of a ~380 nm-thick polycrystalline MgB$_2$ film with the critical temperature ($T_C$) of 39.3 K were investigated at ~8.5 GHz as a function of temperature. Two coherence peaks were observed in the $\sigma_1$ versus temperature curve at temperatures of ~0.5 $T_C$ and ~0.9 $T_C$, respectively, providing a direct evidence for the two-gap nature of MgB$_2$. The film appeared to have a $\pi$-band gap energy of 1.8 meV. For the MgB$_2$ film ion-milled down to the thickness of ~320 nm, two coherence peaks were still observed with the first conductivity peak at ~0.6 $T_C$. Reduction of $T_C$ by 3 K and reduced normal-state conductivity at $T_C$ were observed along with an enhanced $\pi$-band gap energy of 2.1 meV and a reduced $R_S$ at temperatures below 15 K for the ion-milled film. Calculations based on the gap energies from the weak coupling Bardeen-Cooper-Schrieffer theory and the strong coupling theory suggest that both the $\sigma$–band and the $\pi$-band contribute to $\sigma_1$ of the polycrystalline MgB$_2$ films significantly. Our results are in contrast with the observation of single coherence peak at ~0.6 $T_C$ and dominant role of the $\pi$-band in the microwave conductivity of c-axis oriented MgB$_2$ films as reported by Jin *et al.* [Phys. Rev. Lett. 91, 127006 (2003)]. Variations in the inter-band coupling constants with the level of disorder can account for the different $T_C$ and $\sigma_1$ behavior for the as-grown and ion-milled films.

Our results suggest that enhanced inter-band scattering can improve microwave properties of MgB$_2$ films at low temperatures due to the larger $\pi$-band gap despite the reduction of $T_C$ .

## I. Introduction

Since the discovery of superconductivity in magnesium diboride (MgB$_2$)[1], MgB$_2$ has been known as the first superconductor material with two energy gaps, one in the two-dimensional $\sigma$-band and the other in the three dimensional $\pi$-band with very small inter-band scattering between them[2-4]. Accordingly, MgB$_2$ has shown unique properties that have been unheralded from any superconductors with a single gap[5]. Richness in the observed unique properties of MgB$_2$ may also originate from the relative weight with which each band contributes to the measured physical properties. For instance, both bands are known to play an important role in determining the specific heat of MgB$_2$[6,7], and explaining anomalous temperature–dependent anisotropy of the upper critical field[8,9]. Meanwhile Kim *et al.* reported that the measured penetration depth ($\lambda$) of MgB$_2$ is mainly due to the role of the $\sigma$-band[10].

Recently, other interesting results have been reported with regard to the microwave conductivity of MgB$_2$ by Jin *et al.*[11]. They measured the microwave conductivity of MgB$_2$ and reported that contribution of the $\sigma$-band is insignificant with regard to the real part of the complex microwave conductivity of MgB$_2$ films. According to them, observations of single coherence peak at $T/T_C$ ~ 0.6 and non-existence of any coherence peak near $T_C$ support their arguments, with the peak at $T/T_C$ ~ 0.6 attributed to the $\pi$-band. Jin *et al.*'s results, however, appear inconsistent with those by Kim *et al.*[9], who reported that the measured penetration depth, whose magnitude is directly related with the imaginary part of the microwave complex conductivity, is mostly due to the $\sigma$-band. The fact that the films used by both Jin *et al.* and Kim *et al.*

were prepared by the same researchers using the same growth technique makes this kind of inconsistency with regard to the microwave conductivity of $MgB_2$ much more puzzling.

Here we study the surface resistance $R_S$ and the microwave conductivity of high-quality polycrystalline $MgB_2$ films as a function of the temperature. The $R_S$ is measured by a $TE_{011}$ mode dielectric resonator method, the same kind of measurement method used by Jin et al.[11] with $\lambda$ measured by the mutual inductance measurement method proposed by Claassen et al [12]. This gives the absolute value of $\lambda$ independent of the microwave measurements, in contrast to the analysis of Jin et al. [Ref. 11] which infers $\lambda$ from a fit of the microwave data to a specific model. The $R_S$ of an as-grown $MgB_2$ film is compared with that of the same film with reduced thickness after the ion-milling. A crossover in the temperature dependence of the $R_S$ is observed between the two films as has previously been reported for the two kinds of $MgB_2$ films which were separately prepared[13]. Two prominent peaks are observed from both films with each peak attributable to the role of the two bands in the conductivity of $MgB_2$, which is in contrast with that reported by Jin et al [11]. We argue that $R_S$ and the microwave conductivity of the films can be understood in the context of the two-gap scenario.

## II. Experimental

A $MgB_2$ film ($MgB_2$-21A) with the $T_C$ of ~ 39 K was prepared on 0.5 mm-thick c-cut sapphire by the two-step process where boron films deposited on the substrates are annealed in a Mg vapor environment inside quartz tubes. The dimensions of the $MgB_2$ film are 14 x 14 $mm^2$, with the thickness of ~ 380 nm. More details of the sample preparation have been reported elsewhere[14]. The film appeared to be single phase from the X-ray diffraction (XRD) data. Typical dc resistance data also showed that the films had the critical onset temperature of ~ 39 K and the transition width less than ~ 0.3 K.

The effective surface resistance $R_S^{eff}$ of $MgB_2$-21A was measured at temperatures between 6.5 K and 45 K at ~ 8.5 GHz using a $TE_{011}$ mode cavity resonator made of oxygen-free high conductivity copper (OFHC) loaded with a rutile-phase $TiO_2$ (henceforth called 'rutile') rod having the dimensions of 3.88 mm in diameter and 2.73 mm in height. The diameter of the OHFC cavity is 9 mm. During measurements, an epitaxially-grown YBCO film, which was used as a reference, was placed at the bottom of the cavity with the $MgB_2$ film placed at the top. We note that the electric field for the $TE_{011}$ mode is parallel to the surface of each film used as the endplates of the cavity in its direction and has a rotational symmetry around the axis perpendicular to the film surface. The $R_S$ of the YBCO film and the loss tangent of the rutile rod were separately measured to get the $R_S$ of the $MgB_2$ film from the unloaded $Q$ of the rutile-loaded resonator with the $MgB_2$ film at the top and the YBCO film at the bottom. Details for this measurement process have been published elsewhere[15]. The temperature was stable within ± 0.15 K. The $R_S^{eff}$ obtained from the measured unloaded $Q$ was reproducible within 5 % with errors in $R_S^{eff}$ at 8.5 GHz up to ± 15 % below 10 K due to errors in the measured loss tangent of rutile.

The intrinsic surface resistance $R_S$ was calculated using the equation for the effective surface impedance $Z_S^{eff}$. A rigorous $TE_{011}$ mode analysis was performed with the cutoff frequency in the substrate region considered, and the relation between $Z_S^{eff}$ and the intrinsic surface impedance $Z_S$ is expressed by

$$Z_S^{eff} = G_t Z_S = \frac{\coth(\beta_{z3}t) - (\beta_h/\beta_{z3})}{1 - (\beta_h/\beta_{z3})\coth(\beta_{z3}t)} \times \frac{i\omega\mu_0}{\beta_{z3}} \quad (1)$$

Here $Z_S = i\omega\mu_0/\beta_{z3}$ and $\beta_{z3}^2 = i\omega\mu_0\sigma^*$ with $\sigma^*$ ($= \sigma_1 - i\sigma_2$) denoting the microwave complex conductivity, $\omega$, the angular frequency, and $\beta_h = -\beta_{z4}\cot(\beta_{z4}l)$ with $\beta_{z4}$ denoting the complex propagation constant in the substrate region, $t$, the film thickness and $l$, the substrate thickness[16]. $\beta_h$ could be obtained with the measured surface resistance of the OFHC plate at the back of the substrate taken into account. For reference, $\beta_h$ = (-2.07 x $10^3$ – 4.07 x $10^{-1}$ i)/m with $\beta_{z4}$ = (2.26 x $10^{-5}$ – 6.55 x $10^2$ i) at ~ 10 K for $MgB_2$-21A, which changes just a little at different temperatures with the values of $\beta_h$ = (-2.07 x $10^3$ – 4.32 x $10^{-1}$ i)/m and $\beta_{z4}$ = (2.28 x $10^{-5}$ – 6.53 x $10^2$ i) at ~ 38 K. Since $\beta_{z3} = (\omega\mu_0/4)^{1/2}f_1(\sigma_1, \sigma_2)$ and $Z_S = (\omega\mu_0/4)^{1/2}f_2(\sigma_1, \sigma_2)$ with

$f_1 = [\{(p+\sigma_1)^{1/2} + (p-\sigma_1)^{1/2}\} - i\{(p-\sigma_1)^{1/2} - (p+\sigma_1)^{1/2}\}]/p$,

$f_2 = [\{(p+\sigma_1)^{1/2} - (p-\sigma_1)^{1/2}\} + i\{(p-\sigma_1)^{1/2} + (p+\sigma_1)^{1/2}\}]/p$ (2)

and $p = (\sigma_1^2 + \sigma_2^2)^{1/2}$ [17], we took the $\sigma^*$ values that give best fit to the measured $R_S^{eff}$ (i.e., the real part of $Z_S^{eff}$) and $\lambda$ (i.e., the real part of $\beta_{z3}$) and obtained the intrinsic surface resistance $R_S$ from $R_S = Re\{(i\mu_0\omega/\sigma^*)^{1/2}\}$. We note that the $\sigma^*$ values obtained here are almost the same as those from the impedance transformation method[18]. The normal-state intrinsic surface resistance $R_{SN}$ is also calculated from the measured normal-state effective values ($R_{SN}^{eff}$) with the finite thickness of the films taken into account and assuming normal skin effect (i.e., $R_{SN} = X_{SN}$ with $X_{SN}$ denoting the normal-state surface reactance) for MgB$_2$ in the normal state. The normal-state conductivity (resistivity) is 1.26 x 10$^7$/Ω-m (7.94 μΩ-cm) at $T_C$.

Later the as-grown MgB$_2$-21A film was ion-milled and designated MgB$_2$-21I, and then re-measured. In preparing the ion-milled film, the surface of the MgB$_2$ film was etched by argon ion-milling under an angle of 70 degree with respect to the film plane. The etching rate was ~10 nm/min with a beam voltage of 500 V and a current density of 0.28 mA/cm$^2$. MgB$_2$-21I has the $T_C$ of 36.3 K with the thickness ~ 320 nm. The normal-state conductivity (resistivity) is 6.20 x 10$^6$/Ω-m (16.1 μΩ-cm) at $T_C$ for MgB$_2$-21I.

### III. Results and Discussion

Figure 1(a) shows the temperature dependence of $R_S^{eff}$ of MgB$_2$-21A and 21I measured at 8.5 GHz. It is seen that the $R_S^{eff}$ values of both films are lower than ~ 10 μΩ at 7 K, a value comparable to the surface resistance of high-quality MgB$_2$ films including c-axis oriented films[19-21]. This shows that the surface resistance of our film is affected little by their polycrystalline nature at low temperatures. In the figure we see that the surface resistance of MgB$_2$-21I appears different from that of MgB$_2$-21A with increase of ~ 290 mΩ in $R_{SN}^{eff}$ and decrease of ~ 3 K in $T_C$ with $T_C$ of 36.3 K for MgB$_2$-21I. A crossover in the $R_S^{eff}$ vs. $T$ data is also seen between 21A and 21I in the figure, with the $R_S^{eff}$ of 21I appearing lower than that of 21A at temperatures below ~ 13 K. Similar changes in the $R_S^{eff}$ values have repeatedly been observed for high-quality MgB$_2$ films after the ion-milling, which could be explained within the scenario of the two-gap model with the changes in crossover attributed to the enhanced inter-band scattering between the σ-band and the π-band[13]. λ was also measured as a function of temperature for both MgB$_2$ films, as shown in the inset of Fig. 1(a). In the inset, we see that the λ value at 0 K ($λ_0$) are ~ 90 nm for MgB$_2$-21A and ~ 100 nm for MgB$_2$-21I, respectively, which are comparable to the value for c-axis oriented MgB$_2$ films reported by other researchers[11].

We obtained the intrinsic surface resistance $R_S$ of the two MgB$_2$ films from the measured $R_S^{eff}$ and λ values, which are displayed in Fig. 1(b). Here we also see that the surface resistance of MgB$_2$-21I becomes different from that of MgB$_2$-21A with increase of ~ 20 mΩ in $R_{SN}$ and decrease of ~ 3 K in $T_C$ due to the ion-milling. A crossover in the $R_S$ vs. $T$ data is also seen between 21A and 21I in the figure, with the $R_S$ of 21I appearing lower than that of 21A at temperatures below ~15 K. The $R_S$ of both films showed exponential temperature dependence at low temperatures as displayed in Fig. 1(c), which reflects the s-wave nature of the order parameter in the $\pi$-band considering that the $\pi$-band would be mostly responsible for the microwave loss due to quasi-particle excitations at low temperatures. The $\pi$-band gap energies could be obtained from the $R_S$ values of the films at low temperatures, which are ~ 1.8 meV and ~ 2.1 meV for MgB$_2$-21A and MgB$_2$-21I, respectively.

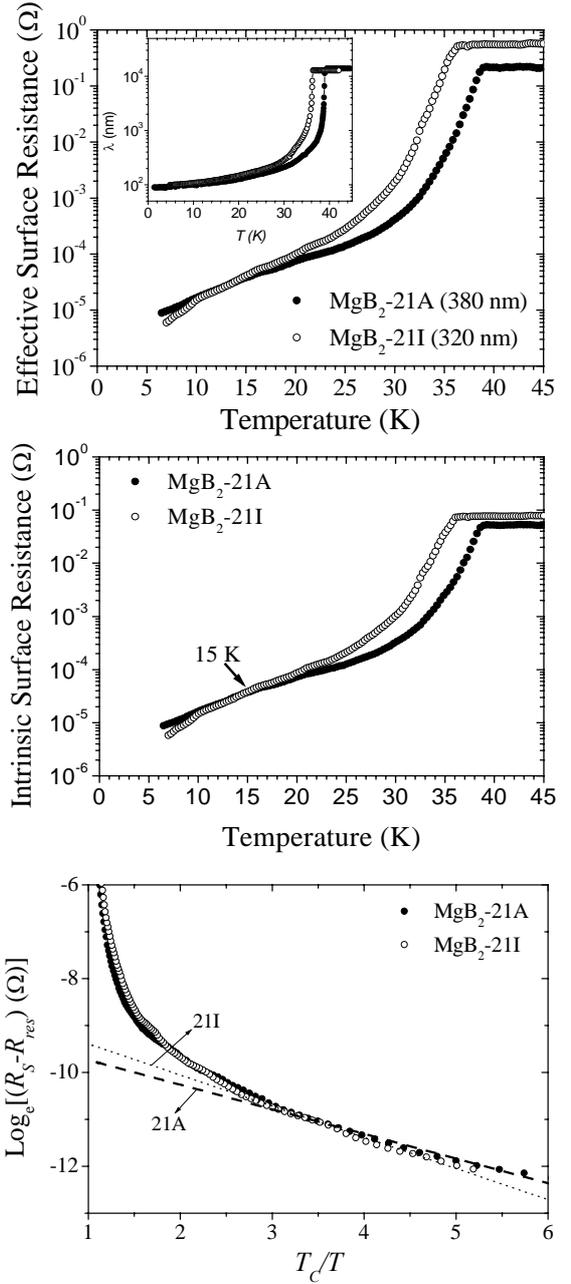

Fig. 1. (a) The effective surface resistance $R_S^{eff}$ of the as-grown MgB$_2$ film MgB$_2$-21A and the ion-milled MgB$_2$ film MgB$_2$-21I. Reduced $T_C$ and Enhanced $R_{SN}^{eff}$ are seen after the ion-milling. A crossover in the $R_S^{eff}$ vs $T$ data is seen at the temperature of ~ 13 K between MgB$_2$-21A and MgB$_2$-21I. Inset: The temperature dependence of the penetration depth for MgB$_2$-21A and MgB$_2$-21I. (b) The intrinsic surface resistance $R_S$ of the as-grown MgB$_2$ film MgB$_2$-21A and the ion-milled MgB$_2$ film MgB$_2$-21I calculated from the measured $R_S^{eff}$ and the penetration depth with the film thickness taken into account. (c) $\ln(R_S - R_{res})$ vs $T_C/T$ curves for MgB$_2$-21A and MgB$_2$-21I. The dashed (dotted) line represents an exponential behavior with the slope characterized by $\Delta(0)/k_BT_C$ = 0.52 (0.66) for MgB$_2$-21A (MgB$_2$-21I). The extrapolated $R_{res}$ is 4 x 10$^{-6}$ μΩ for MgB$_2$-21A, and $R_{res}$ is not subtracted at all for MgB$_2$-21I.

The measured $R_S^{eff}$ and $\lambda$ also enable us to get the values for the conductivity $\sigma_1$ of the MgB$_2$ films, which could be used for investigating effects of its gap energy values on $\sigma_1$. We show the temperature dependence of $\sigma_1$ for the as-grown and the ion-milled MgB$_2$ films in Fig. 2.

We also see two peaks for both films with prominent peaks at ~21 K ($T/T_C$ ~ 0.53) and ~ 36-37 K ($T/T_C$ ~ 0.9) for 21A and the peaks at ~ 22 K ($T/T_C$ ~ 0.61) and ~ 33 K ($T/T_C$ ~ 0.9) for 21I. It seems that the two peaks provide a signature for the existence of two energy gaps in MgB$_2$, with the peaks at $T/T_C$ ~ 0.5 - 0.6 due to the gap in the $\pi$-band and the other peaks at $T/T_C$ ~ 0.9 due to the gap in the $\sigma$-band. This is, however, in contrast with what has been reported by Jin et al., who observed only one peak at $T/T_C$ ~ 0.6 in the $\sigma_1$ vs. $T$ data and attributed the coherence peak to the $\pi$-band [11]. They reported that lack of the coherence peak at the temperature near $T_C$ is due to insignificant contribution of the $\sigma$-band to the real part $\sigma_1$. We could also observe two conductivity peaks for the ion-milled film 21I, with the temperature dependence of $\sigma_1$ appearing somewhat different from that for the as-grown film 21A. We note in Fig. 2 that the peak at ~21 K appears somewhat suppressed compared to the peak near $T_C$ after the ion-milling.

Another interesting thing to note in the figure is the existence of a crossover between the data for 21A and 21I, which shows that existence of the crossover in the $R_S$ vs. $T$ curves for the two MgB$_2$ films is indeed related with their microwave conductivity data with the $\sigma_1$ values of 21I becoming smaller than those of 21A at ~25 K.

To better understand the role of each band in determining the conductivity of the MgB$_2$ films, we provide a theoretical explanation for the conductivity data using the following relation.

$$\sigma_1/\sigma_{1n} = A_\sigma \sigma_{1\sigma}/\sigma_{n\sigma} + B_\pi \sigma_{1\pi}/\sigma_{n\pi}. \quad (3)$$

Here $\sigma_{1n}$ denote the total microwave conductivity in the normal state, with $\sigma_{1\sigma}$ ($\sigma_{1\pi}$) and $\sigma_{n\sigma}$ ($\sigma_{n\pi}$) for the microwave conductivity in the superconducting and normal state due to the $\sigma$-band ($\pi$-band), respectively. $A_\sigma$ and $B_\pi$ denote the relative weighting factors with $A_\sigma + B_\pi = 1$, which determines the relative contribution of each band to the microwave conductivity. In Eq. (3), contribution of inter-band scattering to the conductivity is not considered, with the total conductivity of MgB$_2$ films simply given by the sum of the microwave conductivity of each band. From the values of the $\sigma_{1n}(T_C) = 1.26 \times 10^7/\Omega$-m for MgB$_2$-21A and $6.20 \times 10^6/\Omega$-m for MgB$_2$-21I, and the relation of $\sigma_{1n}(T_C) \sim \varepsilon_0 \hbar \omega_p^2/\Gamma$ with $\hbar \omega_p = 5.9$ eV for the plasma frequency which is assumed equal in the two bands [22], we get $\Gamma \sim 37$ meV and $\Gamma \sim 75$ meV for the scattering rate in MgB$_2$-21A and MgB$_2$-21I, respectively. Since our MgB$_2$ film is regarded as dirty superconductors with the scattering rate much larger than the gap energy of ~ 7 meV for the $\sigma$-band, we used Mattis-Bardeen theory with $\sigma_{1,i}/\sigma_{ni}$ expressed by [23]

$$\frac{\sigma_{1,i}(\omega)}{\sigma_{n,i}(\omega)} = \frac{1}{2\omega} \int_{-\infty}^{\infty} d\Omega \left[ \tanh \frac{\hbar(\Omega+\omega)}{2k_B T} - \tanh \frac{\hbar\Omega}{2k_B T} \right]$$
$$\times [N_i(\Omega)N_i(\Omega+\omega) + M_i(\Omega)M_i(\Omega+\omega)] \quad (4)$$

where

$$N_i(\Omega) = \text{Re}\left\{ \frac{|\hbar\Omega|}{\sqrt{(\hbar\Omega)^2 - \Delta_i^2}} \right\},$$

$$M_i(\Omega) = \text{Re}\left\{ \frac{\Delta_i \, \text{sgn}(\Omega)}{\sqrt{(\hbar\Omega)^2 - \Delta_i^2}} \right\}, \quad (5)$$

with $\omega$ denoting the measured angular frequency, $\Delta_i$, the gap energy in the $\sigma$-band ($\pi$-band) for $i = 1$ (2), $k_B$, the Boltzmann constant, $T$, the temperature. A BCS-like multiple-gap model proposed by Liu et al. was used to obtain the gap energies corresponding to the $\sigma$-band and the $\pi$-band, which is given by

$$\Delta_i = \sum_j \lambda_{ij} \Delta_j \int_0^{\hbar\omega_C} dE \frac{\tanh \frac{\sqrt{E^2+\Delta_j^2}}{2k_B T}}{\sqrt{E^2+\Delta_j^2}}. \quad (6)$$

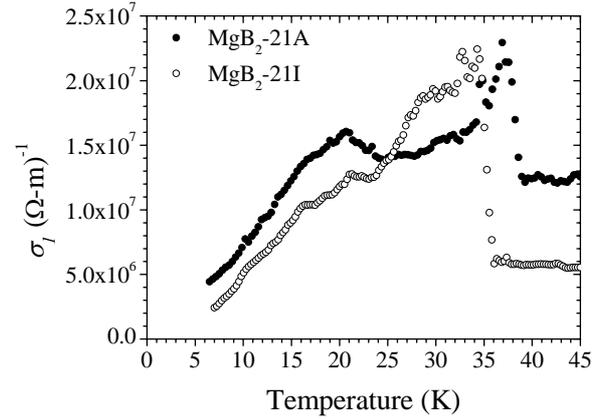

Fig. 2. The temperature dependence of $\sigma_1$ for MgB$_2$-21A and MgB$_2$-21I. Two peaks are shown in the $\sigma_1$ vs. $T$ data at $T/T_C$ ~ 0.53 and ~ 0.9 for MgB$_2$-21A and at $T/T_C$ ~ 0.6 and ~ 0.9 for MgB$_2$-21I. The peak at $T/T_C$ ~ 0.6 for MgB$_2$-21I appears less pronounced compared to that at $T/T_C$ ~ 0.53 for MgB$_2$-21A.

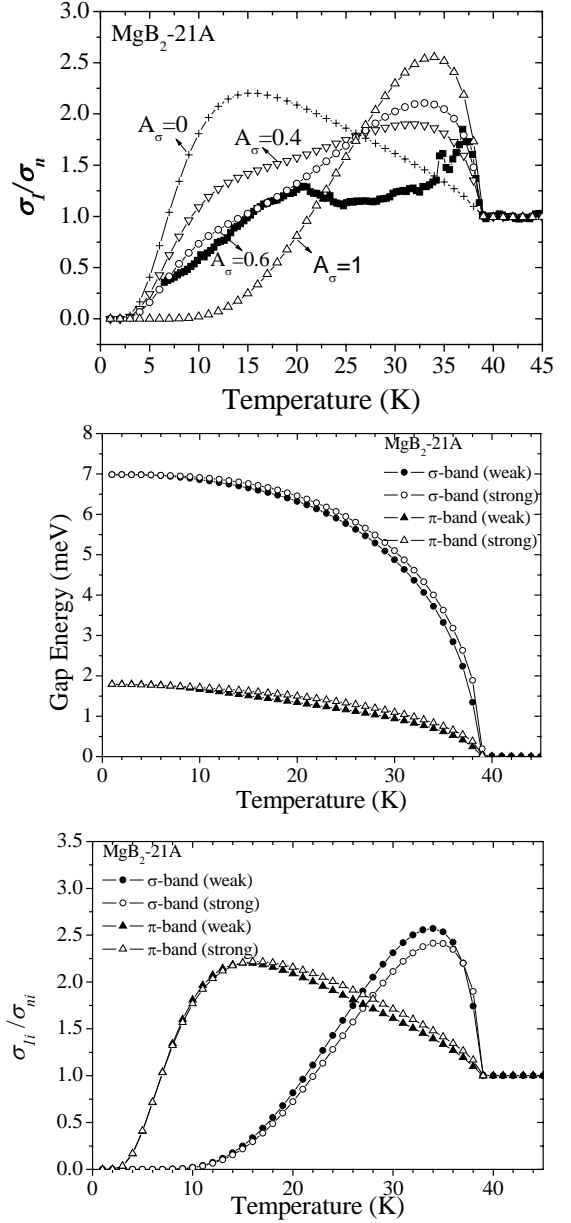

Here $\Delta_i$ is the same as defined in Eq. (4), $\lambda_{ij}$, the electron-phonon coupling constant due to scattering of an electron from band $i$ to band $j$, $\hbar\omega_C$, the cutoff energy which is assumed equal in both bands. We used the values of $\lambda_{11}$ = 0.96, $\lambda_{12}$ = 0.19, $\lambda_{21}$ = 0.14, $\lambda_{22}$ = 0.23 with $\hbar\omega_C$ = 7.5 meV for MgB$_2$-21A having the $T_C$ of 39.3 K and the $\pi$-band gap energy $\Delta_\pi(0)$ of 1.8 meV, where the value for $\lambda_{11}$ is a little smaller than the corresponding value of $\lambda_{11}$ = 0.98 for 40 K MgB$_2$ as given in Ref. 2 with the lower $T_C$ taken into account[24, 25]. In this case, the calculated gap energy values from Eq. (6) with $\hbar\omega_C$ = 7.5 meV become comparable with those by *ab-initio* calculations employing the strong-coupling theory. We note that we used the relation of $\lambda_{ij} = V_{ij}N_j$ in obtaining the $\lambda_{ij}$ values, where $V_{ij}$ denotes the 2 x 2 effective phonon-mediated electron-electron interaction matrix, $N_1$ ($N_2$), the partial density of states of the $\sigma$-band ($\pi$-band) with $N_1 / N_2$ = 1.35 as given in Ref. 2. For reference, the coupling constants used for MgB$_2$-21A correspond to $V_{12}/V_{11} \approx 0.15$. We also note that the difference between the value of $\hbar\omega_C$ = 7.5 meV and the physically relevant logarithmically averaged phonon energy $\hbar\omega_{ln}$ = 56.2 meV could be removed if strong-coupling corrections are considered[2].

In Fig. 3(a), we compare the measured $\sigma_1/\sigma_n$ values with those calculated for different values of $A_\sigma$ and $B_\pi$. In the figure, the measured conductivity appears to agree well with the calculated values for $A_\sigma$ = 0.6 and $B_\pi$ = 0.4 at temperatures below ~ 20 K, showing significant contribution of the $\sigma$-band to the conductivity.

Figure 3(b) shows the gap energies calculated from Eq. (6) using the coupling constants for MgB$_2$-21A, which are compared with those from the strong coupling theory[3]. In the figure, we see that the difference between the two sets of values becomes larger as the temperature gets closer to $T_C$, with the gap energy from the strong-coupling theory larger than that from Eq. (6) by ~ 0.5 meV for the $\sigma$-band and ~ 0.15 meV for the $\pi$-band at 38 K. However, the values of $A_\sigma$ = 0.6 and $B_\pi$ = 0.4 still give the best fit to the measured results even if the gap energies from the strong coupling theory are used. For instance, in Fig. 3(c), we see no significant difference in the overall shapes of the $\sigma_1/\sigma_n$ curve due to the $\sigma$-band ($\sigma_{1\sigma}/\sigma_{n\sigma}$) and that due to the $\pi$-band ($\sigma_{1\pi}/\sigma_{n\pi}$), with the peak position in the $\sigma_{1\sigma}/\sigma_{n\sigma}$ and the $\sigma_{1\pi}/\sigma_{n\pi}$ data remaining the same regardless of the theory used for the two gap energies.

In Fig. 4(a), the experimental conductivity values of

Fig. 3. (a) The measured temperature dependence of $\sigma_1/\sigma_n$ for MgB$_2$-21A compared with the calculated $\sigma_1/\sigma_n$ for different values of $A_\sigma$ and $B_\pi$ in Eq. (3). The best fitted results are seen for $A_\sigma$= 0.6 and $B_\pi$= 0.4 at temperatures below ~ 20 K. (b) A comparison between the calculated gap energy values using Eq. (6) (filled symbols) and those using the strong-coupling theory (open symbols) for the $\sigma$-band and the $\pi$-band of MgB$_2$-21A. The difference becomes larger as the temperature gets closer to $T_C$, with the gap energy from the strong-coupling theory larger than that from Eq. (6) by ~ 0.5 meV and ~ 0.15 meV at 38 K. (c) The temperature dependence of the normalized $\sigma_1$ for the $\sigma$-band ($i$ = 1, circle) and the $\pi$-band ($i$ = 2, triangle) calculated using the gap energy values from Eq. (6) and those from the strong-coupling theory. Changes in the peak position of the curves for the $\sigma$-band and the $\pi$-band appear not significant despite differences in the gap energy values calculated using Eq. (6) and the strong-coupling theory, respectively.

MgB$_2$-21I are compared with those calculated using different values of $A_\sigma$ and $B_\pi$. We used coupling constants of $\lambda_{11} = 0.87$, $\lambda_{12} = 0.245$, $\lambda_{21} = 0.18$, $\lambda_{22} = 0.23$ in calculating the gap energies of MgB$_2$-21I having $T_C$ of 36.3 K and $\Delta_\pi(0)$ of 2.1 meV. In using the coupling constants for MgB$_2$-21I, we considered the following things. At first, to consider the reduction of the $T_C$, which is directly affected by the gap energy in the $\sigma$-band, we used smaller intra-band pairing interaction $V_{11}$ in the $\sigma$-band for MgB$_2$-21I than that for MgB$_2$-21A. Accordingly the intra-band pairing interaction $V_{22}$ in the $\pi$-band for MgB$_2$-21I would be no larger than that for MgB$_2$-21A, and we used the same value for MgB$_2$-21A. Finally, to consider the enhanced $\Delta_\pi(0)$ for MgB$_2$-21I, we used larger inter-band pairing interaction $V_{12}$ (= $V_{21}$) for MgB$_2$-21I than that for MgB$_2$-21A. We could also use smaller $V_{22}$ for MgB$_2$-21I to take the reduced $V_{11}$ into consideration. In this case, however, $V_{12}$ for MgB$_2$-21I should be even larger than the values used here to match the calculated $T_C$ with the measured $T_C$, which doesn't make any significant difference with regard to explaining the reduced $T_C$. For reference, the coupling constants used for MgB$_2$-21I correspond to $V_{12}/V_{11} = 0.21$.

We note that there have been reports on the effects of inter-band scattering on the energy gaps; Brinkman *et al.* predicted collapse of the two gaps into one gap for MgB$_2$ having increased inter-band scattering due to impurities[26] and Yates *et al.* observed correlation between $T_C$ and the electron-phonon coupling constant by studying Raman-active $E_{2g}$ mode of a MgB$_2$ film as a function of thickness through a thin film[27]. In Fig. 4(a), the measured $\sigma_1/\sigma_n$ values appear to agree well with those calculated with $A_\sigma = 0.4$ and $B_\pi = 0.6$ at temperatures below ~25 K, which also suggest significant contribution of the $\sigma$-band to the microwave conductivity of polycrystalline MgB$_2$ films. As mentioned earlier, for the ion-milled sample 21I, the peak at ~22 K appear much suppressed compared to that at 33-34 K, which is different from what we observed for the as-grown sample 21A. We note that the values of $A_\sigma = 0.4$ and $B_\pi = 0.6$ change little even if the gap energies from the strong coupling theory were used. The gap energy values from the weak coupling theory are compared with those by the strong coupling theory in the inset of Fig. 4(a), where the gap energy from the strong coupling theory also gets larger than that from the weak coupling theory at temperatures closer to $T_C$.

We compare $\sigma_1/\sigma_n$ vs. $T/T_C$ data for 21I with those for 21A in Fig. 4(b). One thing we note in the figure is the shift in the peak position related with the $\pi$-band from $T/T_C = 0.53$ for 21A to $T/T_C = 0.61$ for 21I after the ion-milling, which could be attributed to the increased inter-band scattering rate, i.e., the increased inter-band coupling constants for 21I as seen in the inset of Fig. 4(b).

In the inset, the peak position of $T/T_C \sim 0.46$ in the calculated $\sigma_{1\pi}/\sigma_{n\pi}$ for the $\pi$-band of MgB$_2$-21A (labeled 'A') appears shifted to $T/T_C \sim 0.55$ for MgB$_2$-21I (labeled 'I'), which is in qualitative agreements with the change in the peak position due to the $\pi$-band as seen in Fig. 4(a). Meanwhile, we also see in the inset that the peak position of $T/T_C \sim 0.87$ in the calculated $\sigma_{1\sigma}/\sigma_{n\sigma}$ for the $\sigma$-band remains almost the same for MgB$_2$-21A and MgB$_2$-21I despite changes in the coupling constants.

The difference in the microwave conductivity and $T_C$

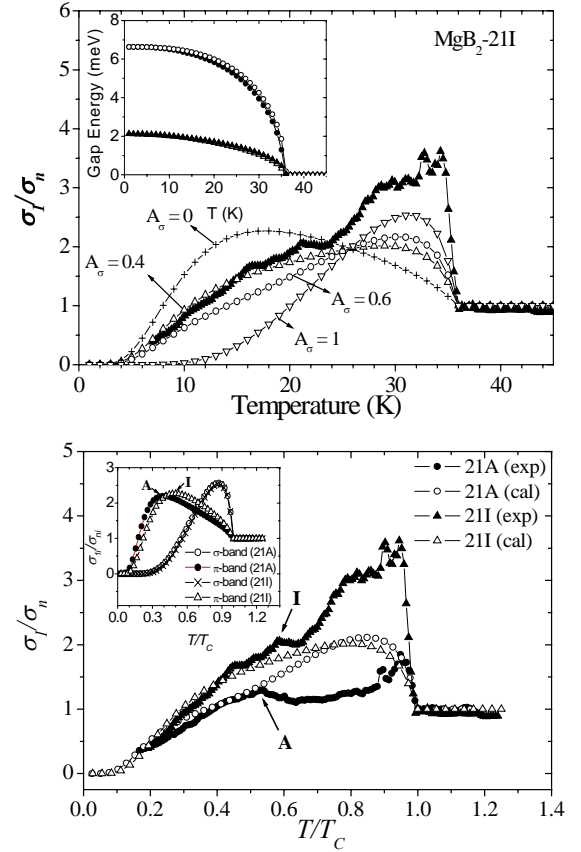

Fig. 4. (a) The measured temperature dependence of $\sigma_1/\sigma_n$ for MgB$_2$-21I compared with the calculated $\sigma_1/\sigma_n$ for different values of $A_\sigma$ and $B_\pi$ in Eq. (3). The best fitted results are seen for $A_\sigma = 0.4$ and $B_\pi = 0.6$ at temperatures below 25 K. Inset: A comparison between the calculated gap energy values using Eq. (6) (filled symbols) and those using the strong-coupling theory (open symbols) for the $\sigma$-band and the $\pi$-band of MgB$_2$-21I. The difference between the values appears smaller than that for MgB$_2$-21A throughout the measured temperature range. (b) The normalized $\sigma_1$ as a function of the reduced temperature $T/T_C$ for MgB$_2$-21A and MgB$_2$-21I. Both the measured (filled symbols) values and the fitted ones (open symbols) are shown. The peak in the $\sigma_1/\sigma_n$ curve at $T/T_C \sim 0.53$ for MgB$_2$-21A (labeled 'A') appears shifted to the right with that at $T/T_C \sim 0.6$ for MgB$_2$-21I (labeled 'I') while no noticeable change is seen for the peak at $T/T_C \sim 0.9$. Inset: The temperature dependence of the normalized $\sigma_{1i}$ ($\sigma_{1i}/\sigma_{ni}$) for the $\sigma$-band ($i = 1$, circle) and the $\pi$-band ($i = 2$, triangle) calculated using the coupling constants for MgB$_2$-21A and those for MgB$_2$-21I.

between MgB$_2$-21A and MgB$_2$-21I might be due to disorders introduced in the MgB$_2$ film during the ion-milling process and/or attributed to the difference in magnesium stoichiometry and oxygen alloying into the boron layers through the thickness of the MgB$_2$ film as pointed out by Yates et al.[27] and Eom et al.[28] An earlier theoretical work shows that enhanced inter-band scattering in a two-gap model of superconductivity indeed increases the smaller gap value despite the reduction of $T_C$ [29].

The difference in the microwave conductivity data between ours and Jin et al.'s[11] cannot be clearly understood at the moment. Since the crystal orientation of our polycrystalline samples is different from Jin et al.'s having c-axis orientations, it is tempting to attribute the observed discrepancy to the difference in the orientations of the MgB$_2$ grains relative to the direction of the measuring currents, considering the nature of the $\sigma$-band having a two-dimensional symmetry. Here it is noted that the microwave conductivity was measured with circulating currents flowing on the surface of the MgB$_2$ films because only the electric field component $E_\phi$ is available in both experiments employing the TE$_{011}$ mode resonator method. However, the difference in the crystal orientations doesn't seem to provide the correct reason considering that the direction of the electric field is parallel to the a-b plane when the TE$_{011}$ mode was used for investigating the properties of c-axis-oriented MgB$_2$ film. In other words, the peak near the $T_C$ due to the $\sigma$-band would be better observed from a sample having c-axis oriented grains, as has been reported in a theoretical work by Dolgov[30], if the scattering rate in the $\sigma$-band is not extremely different from that in the $\pi$-band. There is also a recent report by Gennaro et al. on an experimental observation of the coherence peak near $T_C$ for polycrystalline MgB$_2$[31]. Further studies are needed to address the correct reasons for this.

We note that, considering the three-dimensional nature of the $\pi$-band gap and the strong inter-grain coupling among the grains, the $\Delta_\pi(0)$ values of our films are believed to be reflect the intrinsic properties of our films despite that they are polycrystalline. Since the $\sigma_1$ values due to the $\pi$-band with the smaller energy gap are much larger than those due to the $\sigma$-band at low temperatures, this might explain why good agreements were found between the measured and the calculated $\sigma_1$ at low temperatures below 20 – 25 K (see e.g. Fig. 3(a) and the inset of Fig. 4(b) for the $\sigma_1$ values solely due to the $\pi$-band ($A_\sigma = 0$) and the $\sigma$-band ($A_\sigma = 1$)). At temperatures above 20 – 25 K where contribution of the $\sigma$-band to $\sigma_1$ becomes dominant, it seems necessary to take the relative orientation of the MgB$_2$ grains with respect to the direction of the applied electric field into account in calculating the $\sigma_1$ values, considering the two-dimensional nature of the $\sigma$-band.

Our results suggest that microwave properties of MgB$_2$ films with enhanced inter-band scattering could be improved at low temperatures due to the enhanced energy gap in the $\pi$-band despite reduction of the $T_C$ due to the suppressed energy gap in the $\sigma$-band. We note that enhanced microwave properties have previously been observed in the ion-milled MgB$_2$ films not only in their linear microwave properties such as the $R_S$, but also in their nonlinear microwave properties such as the third harmonic intercept as reported by Booth et al.[32]. With the s-wave nature of the order parameters of MgB$_2$ considered, even a small enhancement of the energy gap in the $\pi$-band by controlled introduction of disorders to the material is expected to lead to substantial improvements in the overall microwave properties.

### IV. Conclusions

The temperature dependences of the surface resistance and the microwave conductivity were measured for high-quality polycrystalline MgB$_2$ films at ~8.5 GHz before and after ion-milling. Two coherence peaks were observed at $T/T_C$ ~ 0.5-0.6 and $T/T_C$ ~ 0.9 in the temperature dependence of the microwave conductivity for both films with the peak due to the $\pi$-band appearing at different reduced temperatures of $T/T_C$ ~ 0.53 for the as-grown film and $T/T_C$ ~ 0.61 for the ion-milled film, which is in contrast with an earlier observation of single coherence peak at $T/T_C$ ~ 0.6 and dominant role of the $\pi$-band in the microwave conductivity of c-axis oriented MgB$_2$ films as reported by Jin et al[11]. Reduction of $T_C$ by 3 K and reduced normal-state conductivity at $T_C$ were observed along with a reduced $R_S$ at temperatures below ~ 15 K for the ion-milled film. The $\pi$-band gap energy of the ion-milled MgB$_2$ film also appeared significantly enhanced with the value of 2.1 meV compared to 1.8 meV for the as-grown MgB$_2$ film. The reduced $R_S$ of the ion-milled MgB$_2$ film appeared attributable to the reduced $\sigma_1$. Calculations based on the gap energies from the weak coupling BCS theory suggest that both the $\sigma$-band and the $\pi$-band contribute to $\sigma_1$ of the polycrystalline MgB$_2$ films significantly with similar results obtained using the gap energies from the strong coupling theory. Increased $T/T_C$ at which the coherence peak due to $\pi$-band is found as well as reduction of $T_C$ observed for the ion-milled film could be explained by increased inter-band coupling constants between the $\sigma$-band and the $\pi$-band which is attributable to increased disorder in the ion-milled film.

Our results suggest that enhanced energy gap in the $\pi$-band of MgB$_2$ films due to enhanced inter-band scattering could lead to improved microwave properties at low temperatures despite reduction of the $T_C$.

**Acknowledgments**

The authors would like to thank M. J. Kim for her help in collecting data and K. T. Leong for useful discussions. We also thank Hyoung Joon Choi, Thomas Dahm and Frank Marsiglio for valuable conversations. The work of S. Y. Lee was partially supported by KOSEF under grant no. R01-2001-00038 and by Korean Ministry of Science and Technology. HJH was supported by Korea Research Foundation Grant (KRF-2002-003-C00042).